\documentclass[11pt,twoside,a4paper,cr,final]{cms-tdr}
\def\svnVersion{64946M}

\begin{document}

\hyphenation{had-ron-i-za-tion}
\hyphenation{cal-or-i-me-ter}
\hyphenation{de-vices}

\RCS$Revision: 62698 $
\RCS$HeadURL: svn+ssh://tdahms@svn.cern.ch/reps/tdr2/papers/XXX-08-000/trunk/XXX-08-000.tex $
\RCS$Id: XXX-08-000.tex 62698 2011-06-21 00:28:58Z alverson $

\newcommand{\mumu}{\ensuremath{\mu^+\mu^-}\xspace}
\newcommand{\qqbar}{\ensuremath{\text{q}\overline{\text{q}}}\xspace}
\newcommand{\QQbar}{\ensuremath{\text{Q}\overline{\text{Q}}}\xspace}
\newcommand{\Jpsi}{\ensuremath{J\hspace{-.08em}/\hspace{-.14em}\psi}\xspace}
\newcommand{\psiP}{\ensuremath{\psi\text{(2S)}}\xspace}
\newcommand{\Ups}[1]{\ensuremath{\Upsilon\text{(#1)}}\xspace}

\newcommand{\doubleRatioUpsB}{\ensuremath{\left.(N_{\PgUb}/N_{\PgUa})_{\PbPb}/(N_{\PgUb}/N_{\PgUa})_{\pp}\right.\xspace}}
\newcommand{\doubleRatioUpsC}{\ensuremath{\left.(N_{\PgUc}/N_{\PgUa})_{\PbPb}/(N_{\PgUc}/N_{\PgUa})_{\pp}\right.\xspace}}

\newcommand{\npart}{\ensuremath{N_{\text{part}}}\xspace}
\newcommand{\ncoll}{\ensuremath{N_{\text{coll}}}\xspace}

\newcommand{\raa}{\ensuremath{R_{AA}}\xspace}
\newcommand{\taa}{\ensuremath{T_{AA}}\xspace}

\newcommand{\eq}[1]{~\ref{#1}\xspace}
\newcommand{\fig}[1]{{Fig.~\ref{#1}}\xspace}
\newcommand{\tab}[1]{table~\ref{#1}\xspace}

\newcommand{\pp}{{\ensuremath{\text{pp}}}\xspace}
\newcommand{\ppbar}{{\ensuremath{\text{p}\overline{\text{p}}}}\xspace}
\newcommand{\PbPb}{\ensuremath{\text{PbPb}}\xspace}
\newcommand{\AuAu}{\ensuremath{\text{AuAu}}\xspace}
\newcommand{\pPb}{\ensuremath{\text{pPb}}\xspace}

\newcommand{\sqrts}{\ensuremath{\sqrt{s}}\xspace}
\newcommand{\sqrtsnn}{\ensuremath{\sqrt{s_{NN}}}\xspace}

\cmsNoteHeader{2013-178} 
\title{$\Upsilon$ suppression in PbPb collisions at the LHC}

\author[llr]{Torsten Dahms,
  for the CMS collaboration}

\date{\today}

\abstract{
  The Compact Muon Solenoid (CMS) has measured the suppression of the
  bottomonium states $\Upsilon\text{(1S)}$, $\Upsilon\text{(2S)}$, and
  $\Upsilon\text{(3S)}$ in PbPb collisions at $\sqrt{s_{NN}} =
  2.76\,\text{TeV}$ relative to pp collisions, scaled by the number of
  inelastic nucleon--nucleon collisions. CMS observed a stronger
  suppression for the weaker bound $\Upsilon\text{(2S)}$ and
  $\Upsilon\text{(3S)}$ states than for the ground state
  $\Upsilon\text{(1S)}$. Such ``sequential melting'' has been
  predicted to be a clear signature for the creation of a quark-gluon
  plasma. The suppression of the $\Upsilon\text{(1S)}$ and
  $\Upsilon\text{(2S)}$ has been measured as a function of collision
  centrality for $\Upsilon$ in the rapidity interval $|y|<2.4$ and
  with transverse momentum ($p_{\mathrm{T}}$) down to 0. Furthermore,
  the $p_{\mathrm{T}}$ and rapidity dependence of the
  $\Upsilon\text{(1S)}$ suppression are presented.
}

\hypersetup{%
pdfauthor={CMS Collaboration},%
pdftitle={Upsilon suppression in PbPb collisions at the LHC},%
pdfsubject={CMS},%
pdfkeywords={heavy-ion collisions quarkonium, bottomonium}
}

\maketitle 

\section{Introduction}
\label{sec:intro}

The goal of the SPS, RHIC, and LHC heavy-ion programmes is to validate
the existence and study the properties of the quark-gluon plasma
(QGP), a state of deconfined quarks and gluons. One of its most
striking expected signatures is the suppression of quarkonium
states~\cite{Matsui:1986dk}, both of the charmonium (\Jpsi, \psiP,
$\chi_c$, etc.) and the bottomonium (\Ups{1S,\,2S,\,3S}, $\chi_b$,
etc.) families. The suppression is predicted to occur above the
critical temperature of the medium ($T_c$) and depends on the \QQbar
binding energy~\cite{Digal:2001ue}. Since the \PgUa\ is the most
tightly bound state among all quarkonia, it is expected to be the one
with the highest dissociation temperature. Examples of dissociation
temperatures are given in Ref.~\cite{Mocsy:2007jz}:
$T_{\text{dissoc}}\sim\!1\,T_c$, $1.2\,T_c,$ and $2\,T_c$ for the
\PgUc, \PgUb, and \PgUa, respectively. Similarly, in the charmonium
family the dissociation temperatures are $\leq1\,T_c$ and $1.2\,T_c$
for the \psiP and \Jpsi, respectively. In \PbPb collisions at the LHC,
\PgU\ mesons are produced at high enough rates to become the prime
choice for measuring a possible sequential melting in a QGP: in
contrast to charmonia, they are not subject to b-hadron feed down and
offer sensitivity to a wider temperature range. As cold-nuclear matter
effects~\cite{Vogt:2010aa} and statistical recombination of
thermalized heavy-quarks~\cite{Andronic:2006ky,Grandchamp:2005yw} are
also expected to be smaller than for charmonia, the interpretation of
\PgU\ suppression is less complicated.

In this proceedings, the CMS measurements of the \Ups{1S,\,2S,\,3S}
mesons via their \mumu decays in \pp and \PbPb collisions at $\sqrtsnn
= 2.76\TeV$ are discussed. The results are first presented as a double
ratio of \PgU\ yields:
\begin{linenomath}
  \begin{align}
    \left.(N_{\Ups{nS}}/N_{\PgUa})_{\PbPb}/(N_{\Ups{nS}}/N_{\PgUa})_{\pp}\right..
  \end{align}
\end{linenomath}
Such double ratio has the advantage that common experimental and
theoretical uncertainties cancel. However, since yields are
uncorrected for acceptance and efficiency, any deviation from unity
will reflect a yield modification of only the bottomonia that are
reconstructed in the CMS detector. Furthermore, it only provides
information about the suppression of the relative suppression of the
excited \PgU\ state with respect to the \PgUa\ ground state. Hence,
the results are also presented as nuclear modifications factors
(\raa), based on a comparison to the yield measured in a \pp reference
run at the same \sqrtsnn, scaled by the number of binary collisions
(\ncoll):
\begin{linenomath}
  \begin{align}
    \raa = \frac{\mathcal{L}_{\pp}}{\taa N_{\text{MB}}}\frac{N_{\PbPb} (\Ups{nS})}{N_{\pp} (\Ups{nS})}\cdot \frac{\varepsilon_{\pp}}{\varepsilon_{\PbPb}}\,.
  \end{align}
\end{linenomath}
The measured yields in \PbPb $(N_{\PbPb}(\Ups{nS}))$ and \pp
collisions $(N_{\pp}(\Ups{nS}))$ are corrected by their respective
efficiencies $\varepsilon_{\PbPb}$ and
$\varepsilon_{\pp}$. $\mathcal{L}_{\pp}$ is the integrated luminosity
of the \pp data set, $\taa$ is the nuclear overlap function, which is
equal to \ncoll divided by the elementary nucleon--nucleon cross
section, and $N_{\text{MB}}$ is the number of minimum bias events in
the \PbPb sample. While the \PgUa\ results as a function of \pt and
rapidity~\cite{Chatrchyan:2012np} are based on the 2010 sample,
corresponding to an integrated luminosity of $\mathcal{L}_{\text
  {int}} = 7.28\mubinv$, the centrality dependent
results~\cite{Chatrchyan:2012lxa} were obtained from the twenty times
larger 2011 data set with an integrated luminosity of
$\mathcal{L}_{\text {int}} = 150\mubinv$. The \pp reference, used in
all results, has an integrated luminosity of $\mathcal{L}_{\text
  {int}} = 231\nbinv$, which for hard-scattering processes is
comparable in size to the 2010 \PbPb sample ($7.28\mubinv \cdot 208^2
\approx 315\nbinv$).

The central feature of CMS is a superconducting solenoid, of 6\,m
internal diameter, providing a field of 3.8\,T. Within the field
volume are the silicon pixel and strip tracker, the crystal
electromagnetic calorimeter (ECAL) and the brass/scintillator hadron
calorimeter (HCAL). Muons are measured in gas-ionization detectors
embedded in the steel return yoke. In addition to the barrel and
endcap detectors, CMS has extensive forward calorimetry. The muons are
measured in the pseudorapidity window $|\eta|< 2.4$, with detection
planes made of three technologies: Drift Tubes, Cathode Strip
Chambers, and Resistive Plate Chambers. Matching the muons to the
tracks measured in the silicon tracker results in a transverse
momentum resolution better than 1.5\% for \pt smaller than 100\GeVc. A
much more detailed description of CMS can be found
elsewhere~\cite{Chatrchyan:2008aa}.

\section{Signal}
\label{sec:signal}

\begin{figure}[ht]
  \begin{center}
    \includegraphics[width=0.4\linewidth]{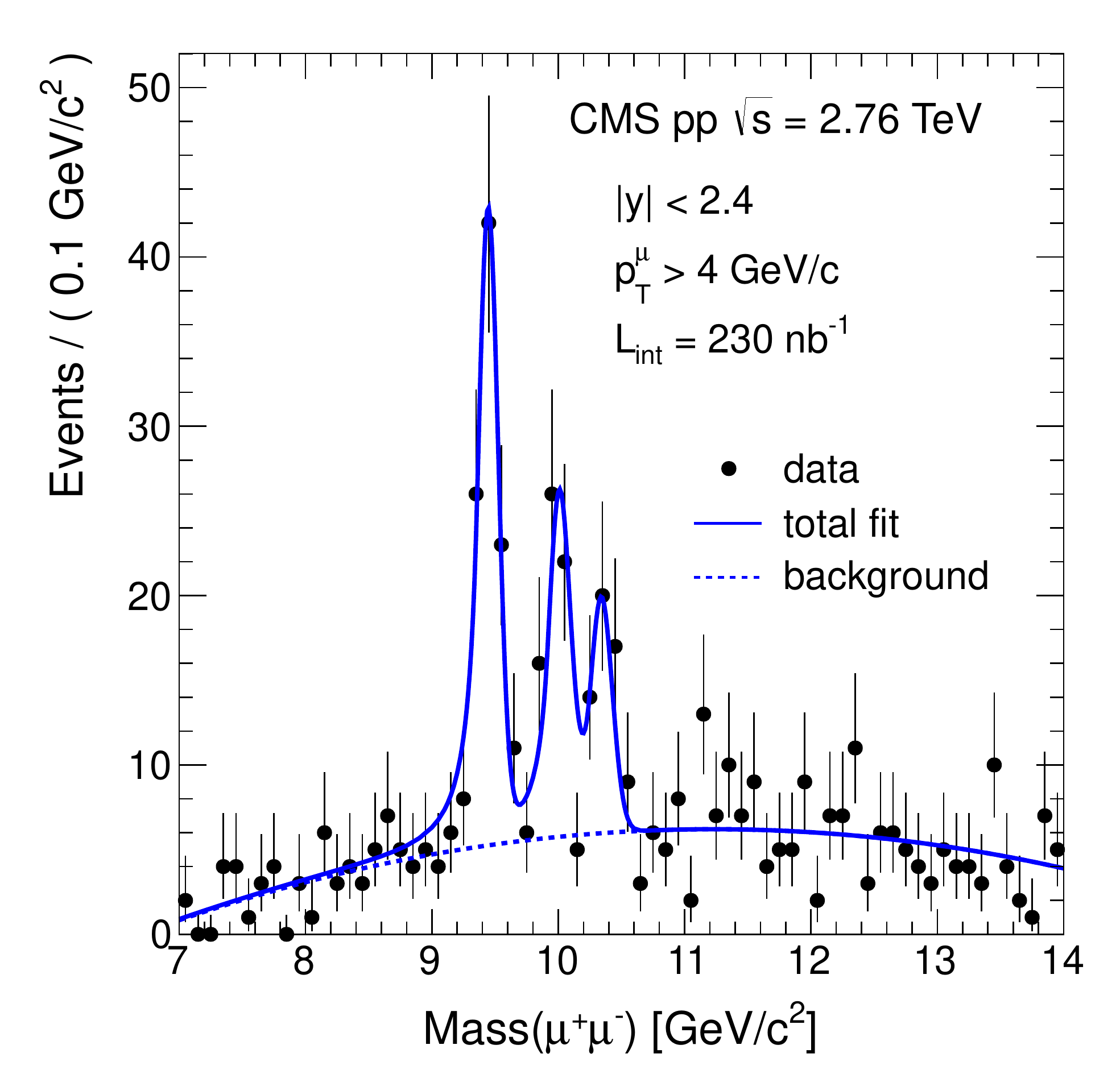}
    \includegraphics[width=0.4\linewidth]{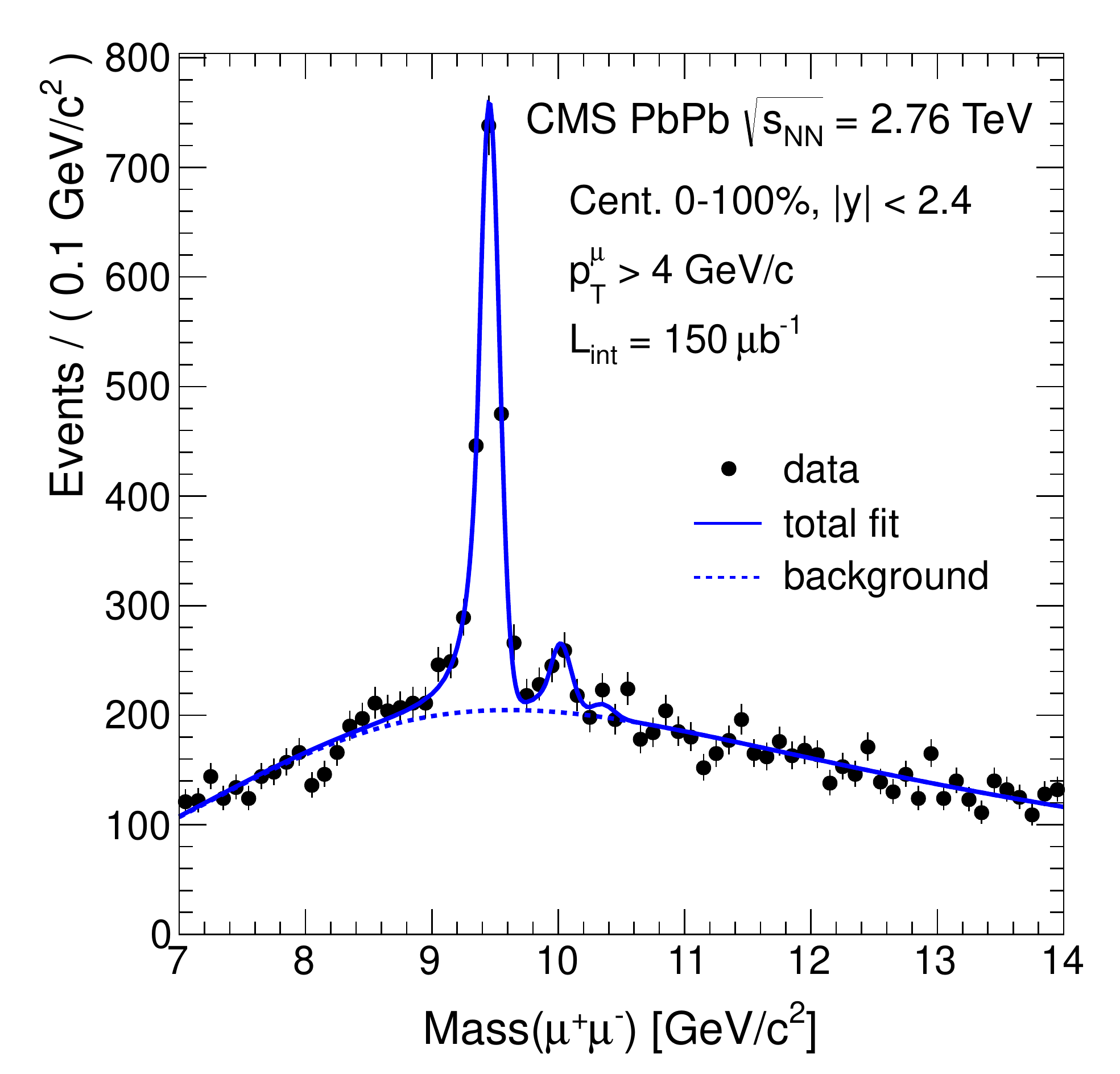}
    \caption{Invariant-mass spectrum of \mumu pairs (black circles) in
      \pp (left) and \PbPb collisions (right) at $\sqrtsnn =
      2.76\TeV$. The solid (signal + background) and dashed
      (background-only) curves show the results of the simultaneous
      fit to the two data sets.}
    \label{fig:invmass}
  \end{center}
\end{figure}

In \fig{fig:invmass}, the invariant-mass spectra of \mumu pairs in the
\PgU\ mass region are shown: the left panel displays the spectrum
measured in \pp collisions, the right panel the spectrum in \PbPb
collisions. Both samples have been reconstructed with the same
algorithm. A transverse-momentum cut of $\pt>4\GeVc$ has been applied
to the individual muons, which still allows \PgU\ with $\pt=0$ to be
reconstructed. Only muons with pseudorapidity $|\eta|<2.4$ have been
reconstructed, restricting the \PgU\ rapidity range to $|y|<2.4$.

The \pp data exhibit clear signals of all three \PgU\ states at their
expected masses. Also in the \PbPb data a clear signal of the \PgUa\
and the \PgUb\ can be seen. However, no clear \PgUc\ peak is
visible. Overlaid is the result of a simultaneous fit to \pp and \PbPb
data. The signal is parametrised by the sum of three Crystal-Ball
functions. A Crystal-Ball function combines a Gaussian core and a
power-law tail at the low side, to account for energy loss due to
final-state photon radiation. The mean and the width of the
Crystal-Ball function, used to describe the \PgUa, are left free in
the fit, while the mean and the width of the \PgUb\ and \PgUc\ signal
shapes are fixed to scale as the mass ratios of the world
average~\cite{Beringer:1900zz}. The \pp background is described by a
parabola, whereas the \PbPb background is fitted to the product of an
exponential function and an error function, which describes the
low-mass turn-on.

\section{Results}
\label{sec:results}

\begin{figure}[ht]
  \begin{center}
    \includegraphics[width=0.32\linewidth]{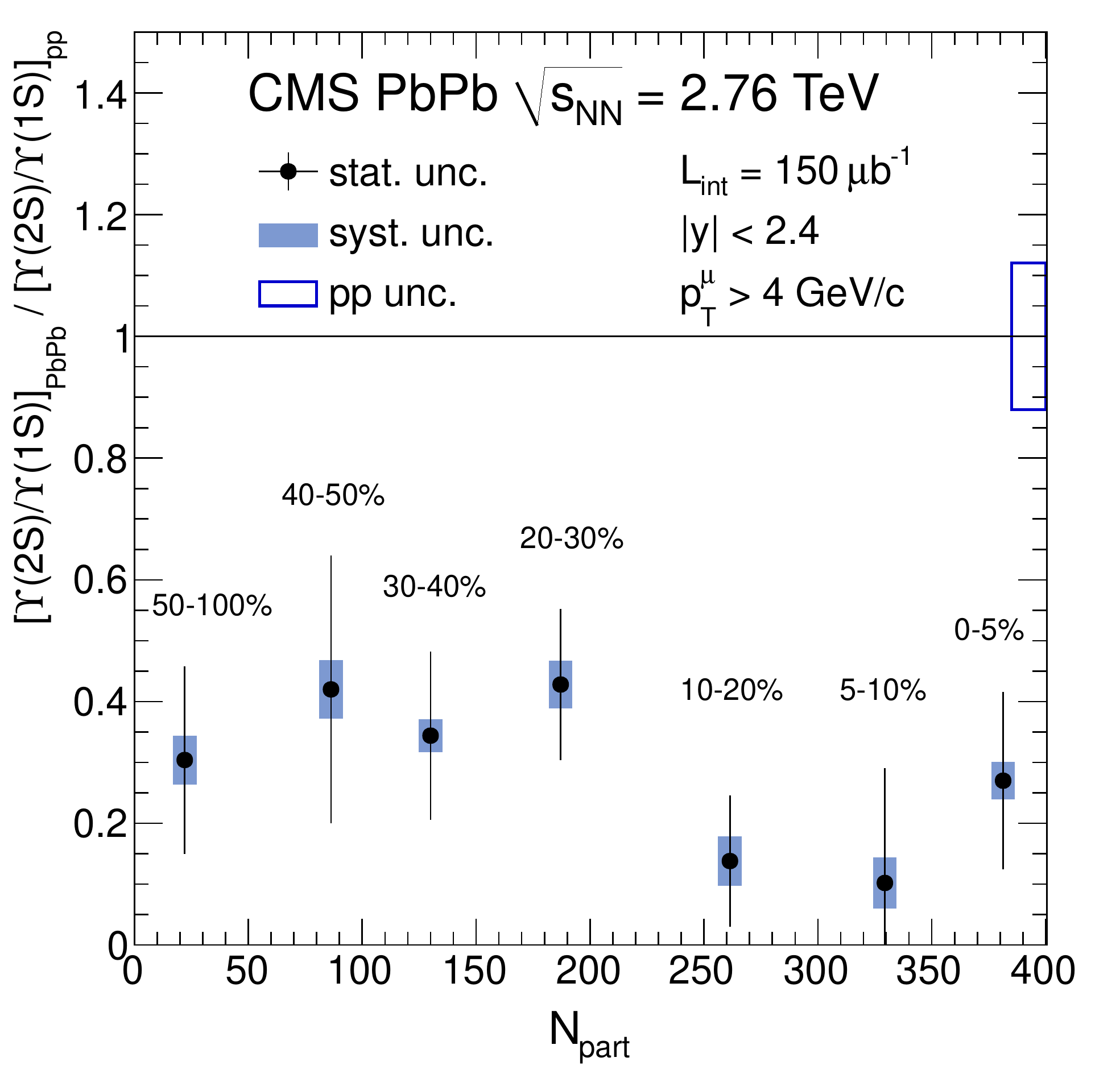}
    \includegraphics[width=0.32\linewidth]{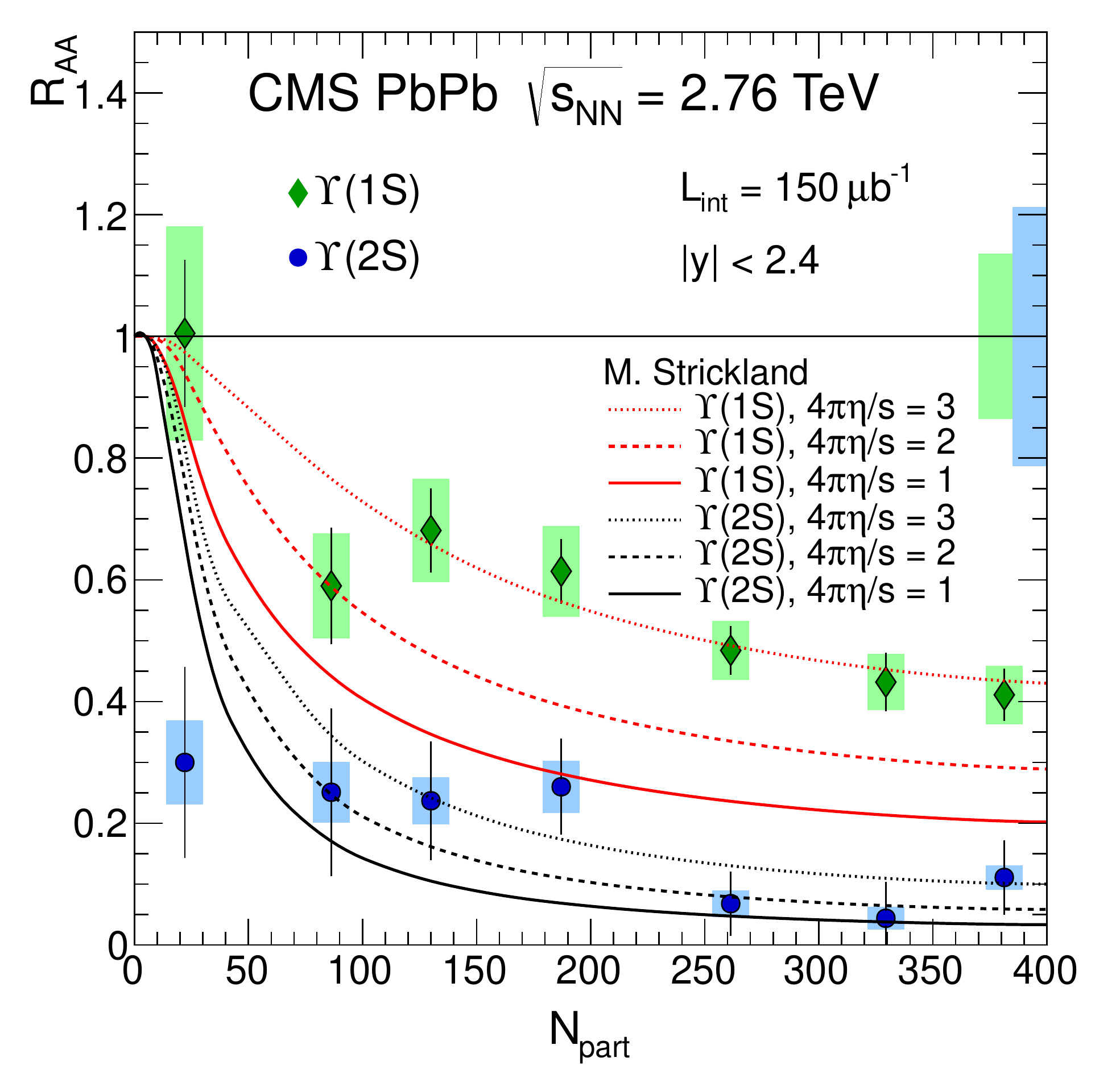}
    \includegraphics[width=0.32\linewidth]{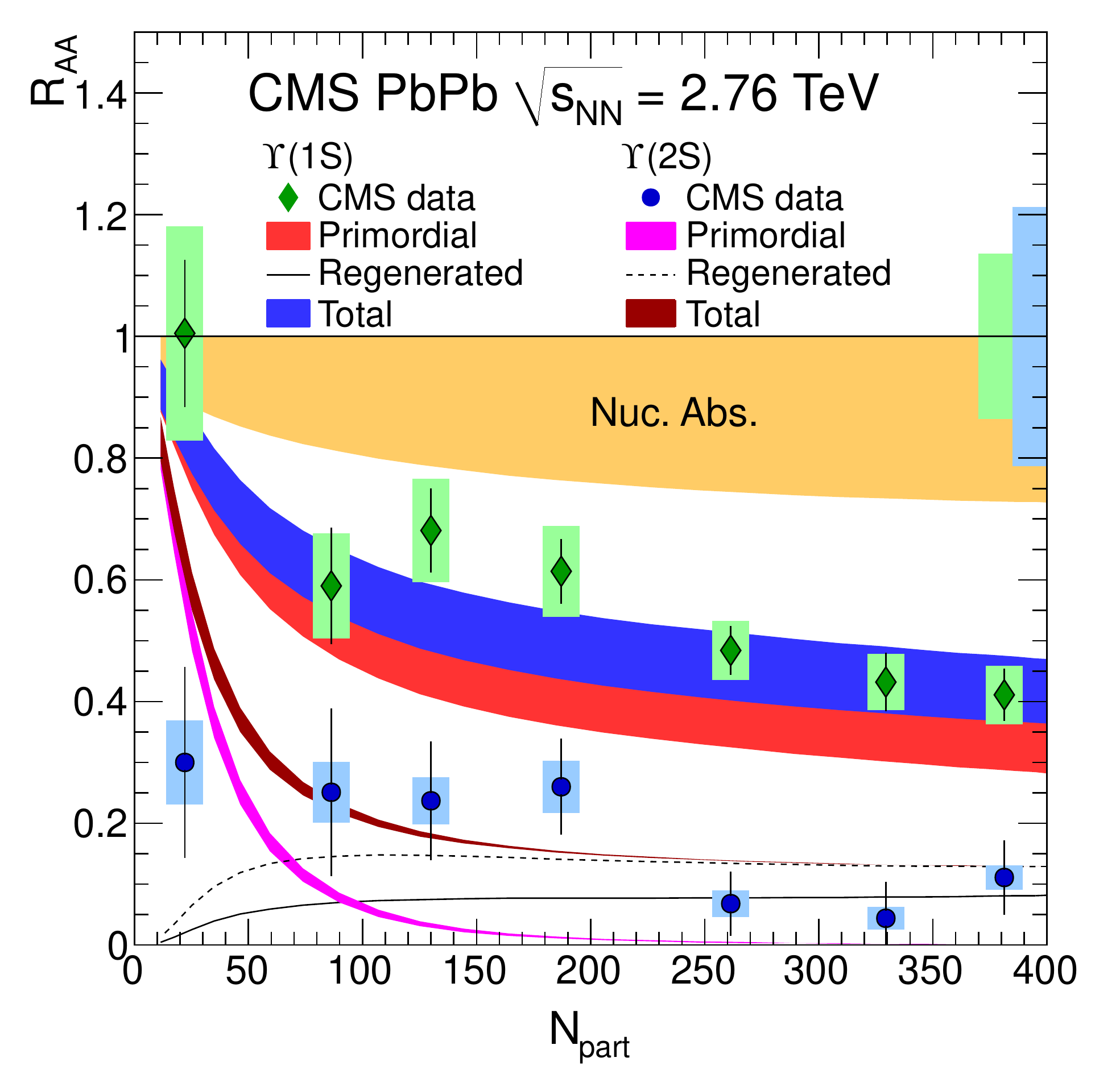}
    \caption{Centrality dependence of the double ratio
      \doubleRatioUpsB (left) and the nuclear modification factor of
      \PgUa\ (green diamonds) and \PgUb\ (blue circles). The data are
      compared to two theoretical models~\cite{Strickland:2012cq}
      (centre) and~\cite{Emerick:2011xu} (right). Statistical
      (systematic) uncertainties are shown as bars (boxes). Global
      uncertainties from the \pp yields and, in case of the \raa,
      luminosity, are shown as boxes at unity.}
    \label{fig:upsilon}
  \end{center}
\end{figure}

From the simultaneous fit to \pp and \PbPb data, centrality integrated
double ratios have been measured separately for the \PgUb\ and \PgUc\
states:
\begin{linenomath}
  \begin{align}
    \doubleRatioUpsB &= 0.21 \pm 0.07\,\text{(stat.)} \pm 0.02\,\text{(syst.)},\notag\\
    \doubleRatioUpsC &= 0.06 \pm 0.06\,\text{(stat.)} \pm 0.06\,\text{(syst.)}\quad(<0.17\text{ at 95\% CL}).\notag
  \end{align}
\end{linenomath}
These double ratios are expected to be unity in the absence of
suppression. Instead, the measured values are significantly
smaller. The \PgUb\ double ratio has been measured as a function of
centrality, as shown in \fig{fig:upsilon}. Within uncertainties, no
pronounced centrality dependence is observed. Furthermore, the nuclear
modification factors of all three states have been measured,
integrated over centrality:
\begin{linenomath}
  \begin{align}
    \raa(\PgUa) &= 0.56 \pm 0.08\,\text{(stat.)} \pm 0.07\,\text{(syst.)},\notag\\
    \raa(\PgUb) &= 0.12 \pm 0.04\,\text{(stat.)} \pm 0.02\,\text{(syst.)},\notag\\
    \raa(\PgUc) &= 0.03 \pm 0.04\,\text{(stat.)} \pm 0.01\,\text{(syst.)}\quad(<0.10\text{ at 95\% CL}).\notag
  \end{align}
\end{linenomath}
The centrality dependence of the nuclear modification factors of
\PgUa\ and \PgUb\ are displayed in the centre panel of
\fig{fig:upsilon}. The data show a clear ordering of the suppression
with binding energy, the least bound state being the most
suppressed. The suppression of the \PgUa\ state is consistent with no
suppression of directly produced \PgUa, but a suppression of feed-down
contribution from excited state decays only, which is expected to
contribute $\approx 50\%$ at high \pt~\cite{Affolder:1999wm}. However,
the uncertainties in the measurement of the feed-down contributions
preclude quantitative conclusions about the suppression of directly
produced \PgUa.

These results are compared to theoretical models of \PgU\ suppression
in a quark-gluon plasma. The first, presented in the centre panel of
\fig{fig:upsilon}, combines a complex-valued heavy-quark potential and
a dynamic evolution of an anisotropic
plasma~\cite{Strickland:2012cq}. The model is presented for three
values of shear viscosity to entropy density ratios $\eta/s$. To fix
the charged particle multiplicity, for each value of $\eta/s = \{1, 2,
3\}$ a different initial temperatures is assumed: $T_0 = \{520, 504,
494\}\MeV$. The \PgUa\ data appear to favour larger values
$\eta/s$. The second model, compared to CMS data in the right panel of
\fig{fig:upsilon}, is based on a rate-equation
approach~\cite{Emerick:2011xu}. It is worth to note that this model
predicts sizeable cold nuclear matter effects on \PgU\ states produced
in \PbPb collisions at the level of $\approx 20\%$. The authors also
predict a small contribution of \PgU\ production via regeneration of
thermalized b quarks, which provides the only source of \PgUb\ in
central \PbPb collisions. While both models describe the data
qualitatively, for a detailed comparison it is crucial to not only
decrease the global uncertainty on the \raa measurement, which is
currently limited by the available \pp data. But it is also mandatory
to measure the feed-down contributions with better precision and to
low \pt, as these enter directly in the model calculations.

\begin{figure}[ht]
  \begin{center}
    \includegraphics[width=0.4\linewidth]{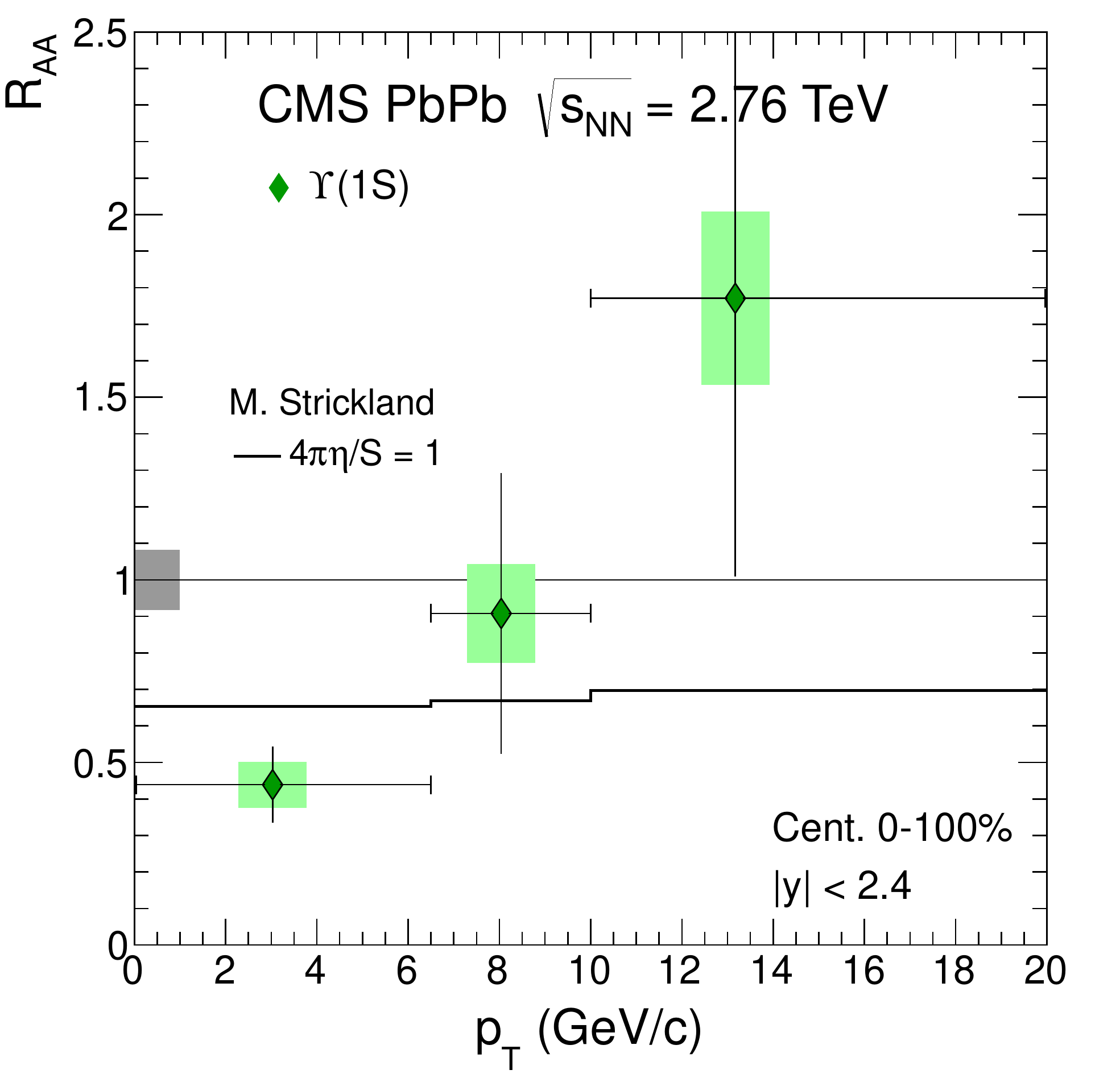}
    \includegraphics[width=0.4\linewidth]{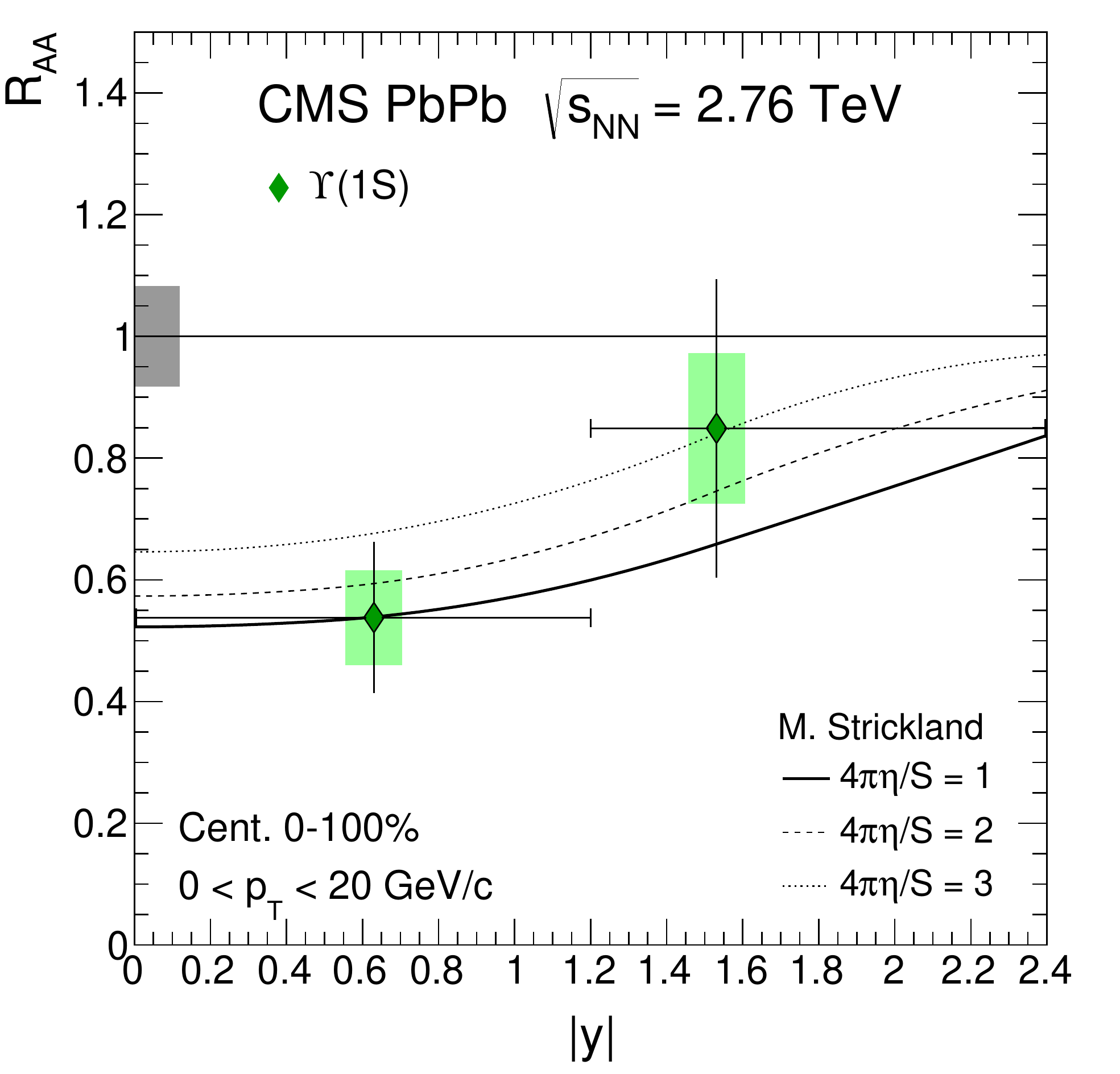}
    \caption{Transverse momentum (left) and rapidity (right)
      dependence of the nuclear modification factor of \PgUa\ (green
      diamonds). The data are compared to a theoretical
      model~\cite{Strickland:2011mw}. Statistical (systematic)
      uncertainties are shown as bars (boxes). Global uncertainties
      from the \pp yields, \taa, and luminosity are shown as boxes at
      unity.}
    \label{fig:upsPtRap}
  \end{center}
\end{figure}

A first attempt to measure the \pt and rapidity dependence of the
\PgUa\ suppression has been made with the 2010 \PbPb data, as shown in
\fig{fig:upsPtRap}. At this point it is too early to conclude on a \pt
and rapidity dependence of the \PgUa\ suppression. However, it is
clear that \PgUa\ at low-\pt and midrapidity are suppressed. The data
are compared to an earlier prediction of the same model as in the
bottom-left panel of \fig{fig:upsilon}~\cite{Strickland:2011mw}. The
model follows the general features of the data, but underpredicts the
suppression of \PgUa\ at low \pt.

\section{Summary}
\label{sec:summary}

In summary, CMS has measured the nuclear modification factors of the
\PgUa, \PgUb, and \PgUc\ mesons. For \PgUa\ and \PgUb, these are
presented as a function of centrality. Furthermore, the \raa as a
function of \pt and rapidity has been measured for the \PgUa. A
sequential melting of the \PgU\ states is observed. Models of \PgU\
suppression in a QGP qualitatively describe the data. New data from
\pp collisions at $\sqrts = 2.76\TeV$ and \pPb collisions at $\sqrtsnn
= 5.02\TeV$, collected at the beginning of 2013, will enable CMS to
study the \pt and rapidity dependence of the \PgU\ \raa in more detail
and to quantify cold nuclear matter effects on the double ratio and
the \raa.

\section*{Acknowledgements}
\label{sec:ack}

TD received funding from the European Research Council under the FP7
Grant Agreement no. 259612.

\end{document}